\documentclass[namedreferences,natbib]{solarphysics}
\usepackage[optionalrh,solaenum]{spr-sola-addons} 
\usepackage{graphicx}                    
\usepackage{color}                       
\usepackage{url}                         

\begin{document}

\begin{article}

\begin{opening}

  \title{On the Relation between Solar Activity and Clear-Sky
    Terrestrial Irradiance}

%
\author{G.~\surname{Feulner}
       }

%
\runningauthor{Feulner}
\runningtitle{Solar Activity and Terrestrial Irradiance}

%
\institute{G.\ Feulner\\Earth System Analysis, Potsdam Institute for
  Climate Impact Research (PIK), P.O.~Box~60\,12\,03, D--14412~Potsdam,
  Germany\\e-mail: \url{feulner@pik-potsdam.de}}

\begin{abstract}
  The Mauna Loa Observatory record of direct-beam solar irradiance
  measurements for the years 1958--2010 is analysed to investigate the
  variation of clear-sky terrestrial insolation with solar activity
  over more than four solar cycles. The raw irradiance data exhibit a
  marked seasonal cycle, extended periods of lower irradiance due to
  emissions of volcanic aerosols, and a long-term decrease in
  atmospheric transmission independent of solar activity. After
  correcting for these effects, it is found that clear-sky terrestrial
  irradiance typically varies by $\approx 0.2 \pm 0.1$\% over the course
  of the solar cycle, a change of the same order of magnitude as the
  variations of the total solar irradiance above the atmosphere. An
  investigation of changes in the clear-sky atmospheric transmission
  fails to find a significant trend with sunspot number. Hence there
  is no evidence for a yet unknown effect amplifying variations of
  clear-sky irradiance with solar activity.
\end{abstract}

%

\end{opening}

%

\section{Introduction}
\label{s:intro} 

Long-term records of clear-sky terrestrial solar irradiance are very
useful to study changes in atmospheric transmission
\citep[\textit{e.g.},][]{Hoyt1983}. It is more difficult to derive changes in
insolation due to solar variability from these data because the
variations related to solar activity are much smaller than the changes
in atmospheric transmission of different origin and the systematic and
random errors in the data. Nevertheless, such an approach has been
tried many times in the past.

The Smithsonian Astrophysical Observatory (SAO) observations from the
first half of the 20th century are an important historical example for
terrestrial irradiance measurements with the aim to detect solar
variability from the ground \citep{Abbot1932, Abbot1942,
  Aldrich1954}. The SAO observing campaigns were designed to derive a
time series of the solar constant or total solar irradiance (TSI)
above the atmosphere, but calibration issues \citep{Allen1958}, biases
due to strong seasonal variations (\citealt{Feulner2011b}a), and
changes in atmospheric transmission independent of solar activity
\citep{Angstrom1970} -- mostly related to volcanic aerosols and local
pollution \citep{Hoyt1979, Roosen1984} -- can pose serious
difficulties for investigations of irradiance changes related to solar
activity if not properly accounted for.

Despite these problems, the SAO data continue to attract
attention. \citet{Weber2010, Weber2011} correlated the ground-based
irradiance measurements from the SAO data with sunspot numbers and
claimed to have found a strong variation of terrestrial insolation
with solar activity, with differences of the order of 1\% between
solar maxima and minima (\textit{i.e.}, one order of magnitude larger
than the variations in the TSI on top of the atmosphere). These
findings were criticised by \citeauthor{Feulner2011b} (2011a, b)
\nocite{Feulner2011d} who could show that they were due to seasonal
bias and the effects of volcanic aerosols and local pollution. The
seasonal bias arises from the fact that, by coincidence, days with
large sunspot numbers in the SAO data predominantly lie in months with
seasonally high atmospheric transmission, making irradiance values at
times of high solar activity appear higher. The second bias is
introduced because two out of three solar minima in the SAO data
covering solar cycles 16, 17, and 18 are affected by reduced
atmospheric transmission due to volcanic aerosols and local
pollution. This effect makes solar irradiance on the ground during
those two minima appear lower, thus again suggesting a stronger
variation of terrestrial irradiance with solar activity. After
correcting for the seasonal cycle and excluding periods of time
affected by aerosols from volcanic eruptions and local pollution, the
variations of terrestrial insolation are of the same order of
magnitude as the TSI variations (\citeauthor{Feulner2011b}, 2011a,
b)\nocite{Feulner2011d}.

More recently, \citet{Hempelmann2012} used terrestrial insolation data
taken at Mauna Loa Observatory (MLO) and reported variations of the
solar irradiance on the ground a factor of 10 larger than above the
atmosphere. Here an improved analysis of these data is presented to
independently quantify how much clear-sky terrestrial irradiance
varies between solar maxima and minima.

This paper is organised as follows. Section~\ref{s:data} describes the
Mauna Loa terrestrial insolation data and corrections of the effects
of the seasonal cycle, aerosols from volcanic eruptions, and a linear
long-term trend independent of solar activity. In
Section~\ref{s:correlation} the correlation of the thus corrected
irradiance with sunspot number is presented. Section~\ref{s:atf}
analyses the atmospheric transmission factor and its changes with
sunspot number, before the results are discussed in the context of
previous work in Section~\ref{s:disc}. Finally, Section~\ref{s:concl}
summarises and concludes this paper.

\section{Data Analysis}
\label{s:data}

\subsection{A First Look at the Raw Irradiance Data}

\begin{figure}[ht!]
\centerline{\includegraphics[width=0.7\textwidth]{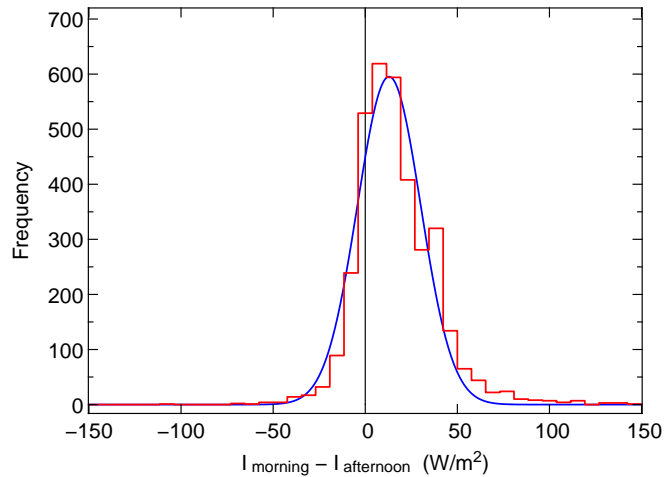}}
\caption{Difference of morning and afternoon irradiance measurements
  at MLO and airmass 2 for days where both are available (red
  histogram), showing that morning values are on average larger than
  afternoon values. The blue curve shows a Gaussian function fit to
  the data with a peak at 12.9~W~m$^{-2}$ and a width $\sigma =
  17.2$9~W~m$^{-2}$.}\label{f:daydiff}
\end{figure}

For the analysis in this paper, the ground-based direct-beam solar
irradiance data taken since 1958 at Mauna Loa Observatory (MLO,
latitude 19.533$^\circ$~N, longitude 155.578$^\circ$~W, elevation
3\,400~m) are used \citep{Ellis1971, Dutton1985, Dutton1994,
  Dutton2001}. These measurements have been obtained at local noon as
well as at airmass values of 2, 3, 4, and 5 (both in the morning and
in the afternoon). In the following, results for the morning
irradiance data at airmass 2 are presented. Airmass 2 is chosen since
smaller airmass values correspond to smaller zenith angles and thus
shorter paths through the atmosphere. Furthermore, as in most previous
studies based on these data, the morning measurements are preferred
over the afternoon data or a combined record of the two due to the
absence of local influences on the atmospheric transmission because of
downslope winds in the morning hours \citep{Mendonca1969}. Indeed, a
comparison of irradiance morning and afternoon data at airmass two
for days where both measurements are available shows that morning
irradiance values are larger on average and thus apparently less
affected by local disturbances in atmospheric transmission (see
Figure~\ref{f:daydiff}).

As a consistency check, the full analysis has been carried out for a
combined record of morning and afternoon data and for the other airmass
values as well, however, yielding very similar results which are not
shown in this paper.

\begin{figure}[ht!]
\centerline{\includegraphics[width=0.95\textwidth]{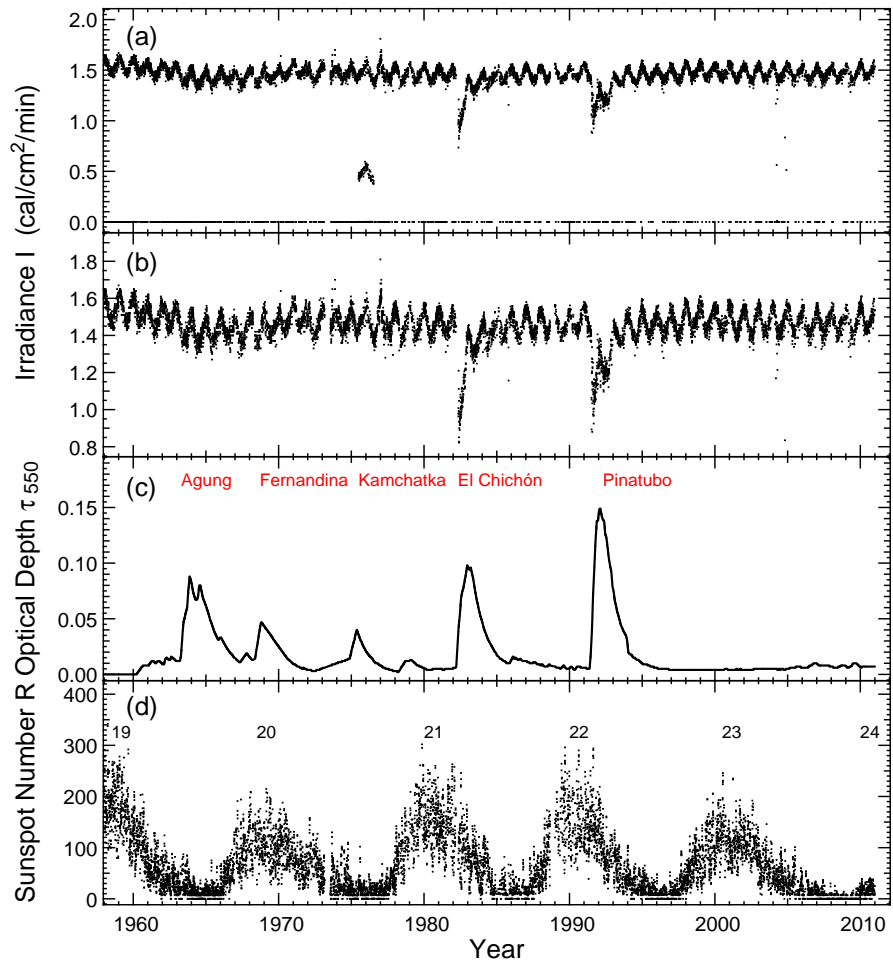}}
\caption{The MLO terrestrial irradiance data for the time period
  1958--2010 in the context of solar activity and volcanic eruptions
  during this time interval. (a) Raw irradiance measurements for
  morning measurements at airmass 2. (b) Same as (a), but zooming into
  the data. (c) Global average optical depth $\tau_{550}$ of volcanic
  aerosols at $\lambda = 550$~nm from \citet{Sato1993} until the end
  of 1993 and from \citet{Solomon2011} since the beginning of
  1994. Prominent volcanic eruptions are marked. (c) Daily
  international sunspot numbers $R$ during this time interval,
  spanning a large part of solar cycle 19, cycles 20, 21, 22, and 23
  as well as the beginning of the current cycle 24.}\label{f:mlotime}
\end{figure}

The raw irradiance data at morning airmass 2, for historic reasons
given in units of Langley~min$^{-1}$ or cal~cm$^{-2}$~min$^{-1}$, are
shown in Figures~\ref{f:mlotime}a and \ref{f:mlotime}b. In this
figure, they are compared to the globally averaged optical depth
$\tau_{550}$ of volcanic aerosols at a wavelength of $\lambda =
550$~nm as compiled by \citet{Sato1993} for the years 1959--1993 and
by \citet{Solomon2011} for 1994--2010 (Figure~\ref{f:mlotime}b) and to
daily international (also called Wolf or Z\"urich) sunspot
numbers\footnote{Available at
  \url{http://sidc.oma.be/DATA/dayssn_import.dat}, date of access: 22
  December 2011.} $R$ (Figure~\ref{f:mlotime}c).

Several interesting observations can be made in the raw data which are
relevant for a proper analysis of correlations with solar activity:

\begin{enumerate} 

\item The irradiance data exhibit a pronounced seasonal cycle. As
  shown in \citeauthor{Feulner2011b} (2011a), not correcting for
  seasonal variations can affect correlations with solar activity if
  different phases of solar activity are not evenly distributed over
  the seasons in the data.

\item There is a cluster of irradiance data around
  0.5~cal~cm$^{-2}$~min$^{-1}$ in the years 1975--1976. As one can
  clearly see the seasonal cycle and as these values are offset by 1
  from the majority of the data; these are most likely due to typos in
  the record where values have a leading digit of 0 instead of
  1. These typos can either be corrected or masked out; the latter
  option has been chosen in this paper.

\item The dimming effects of volcanic aerosols in the atmosphere can
  be seen directly in the data. These are most pronounced for the
  largest eruptions, \textit{i.e.}, El~Chich\'on in 1982 and Pinatubo
  in 1991, of course, but irradiance values are also lower during
  volcanic episodes in the 1960s and 70s. This can be best seen in the
  expanded view presented in Figure~\ref{f:mlotime}b.

\item Looking at periods only weakly affected by volcanic aerosols
  over the entire record, one can see a slight decrease of terrestrial
  irradiance with time previously described in \citet{Solomon2011}. It
  will be shown below that this long-term trend is well approximated
  by a linear fit and not connected to changes in sunspot number.

\end{enumerate}

All these issues have to be taken care of before correlating the MLO
terrestrial irradiance data with sunspot number in order to ensure an
unbiased estimate of the influence of solar activity on clear-sky
irradiance on the ground. In addition, the MLO data are also affected
by changes in instrumentation and offsets in calibration which can
also influence any trend analysis
\citep[\textit{e.g.},][]{Dutton2001}. These effects can be eliminated,
however, by directly investigating atmospheric transmission factors
\citep{Ellis1971} as shown in Section~\ref{s:atf}.

\subsection{Corrections Applied to the Irradiance Data}
\label{s:corrections}

\begin{figure}[ht!]
\centerline{\includegraphics[height=0.95\textwidth,angle=270]{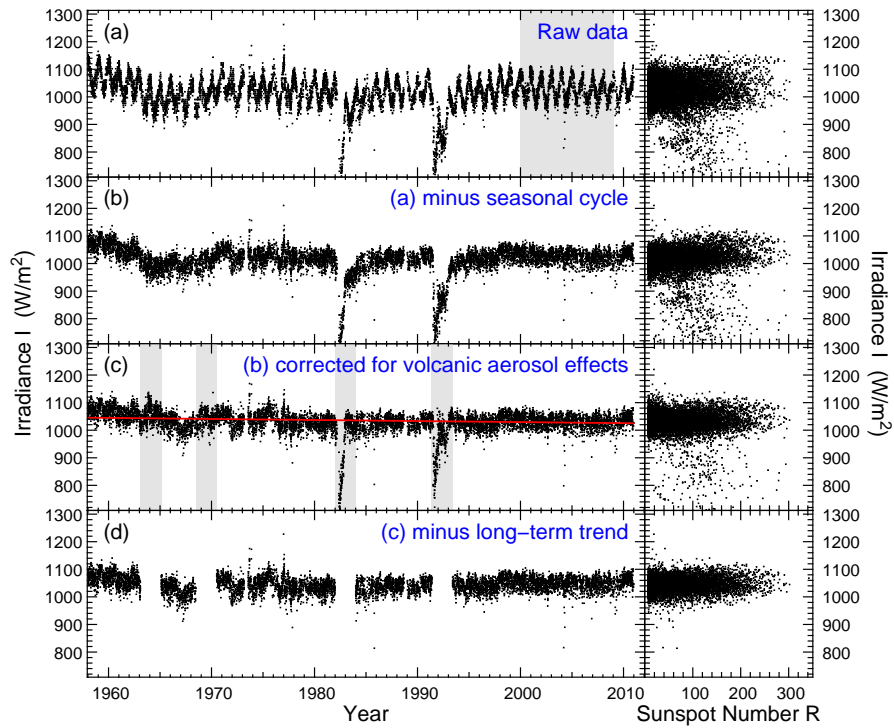}}
\caption{Corrections applied to the MLO terrestrial irradiance data at
  airmass 2. The left-hand panels show the irradiance time series, the
  right-hand panels scatter plots of irradiance vs.\ sunspot
  number. (a) Raw irradiance data as shown in Figures~\ref{f:mlotime}a
  and \ref{f:mlotime}b. The grey shading indicates the portion of the
  data used to derive the average seasonal cycle in terrestrial
  irradiance. (b) The raw data with the seasonal cycle subtracted. (c)
  The data corrected for both seasonal cycle and for the effects of
  volcanic aerosols using the globally averaged optical depth shown in
  Figure~\ref{f:mlotime}b. The two-year time periods with residuals
  after the El~Chich\'on and Pinatubo eruptions excluded in the
  analysis below are indicated by the grey shading. The red line shows
  the linear long-term trend derived from a fit to the data. (d) Same
  as (c), but with the long-term trend
  subtracted.}\label{f:mlotimecorr}
\end{figure}

As a first step, the raw MLO irradiance data for morning airmass 2 as
shown in Figures~\ref{f:mlotime}a and \ref{f:mlotime}b are converted
to SI units (W~m$^{-2}$) by multiplying by a factor of 697.4. These
converted irradiance data are shown both as a time series and as a
scatter plot vs.\ sunspot number in
Figure~\ref{f:mlotimecorr}a. Furthermore, for each day with
observations the MLO irradiance data are correlated with daily
international sunspot numbers and interpolated volcanic aerosol
optical depth data from the same day.

In the discussion above, three sources of systematic bias of the
irradiance data were identified: the seasonal cycle, attenuation by
volcanic aerosols, and a possible long-term trend. These effects will
be discussed and, if possible, corrected in the following. The
individual corrections and their effect on the time series as well as
the correlation between irradiance and sunspot number are shown in
Figure~\ref{f:mlotimecorr}.

\subsubsection{Seasonal Cycle}

The pronounced seasonal cycle is one of the most obvious features of
the raw irradiance time series shown in
Figure~\ref{f:mlotimecorr}a. One way to correct for this effect is to
take data from several years only little affected by volcanic aerosols
and combine them to compute median monthly values for the irradiance
(\citeauthor{Feulner2011b}, 2011a). Daily values of this average
seasonal irradiance cycle can then be computed from a cubic-spline fit
through the data with periodic boundary conditions.

\begin{figure}[ht!]
\centerline{\includegraphics[width=0.7\textwidth]{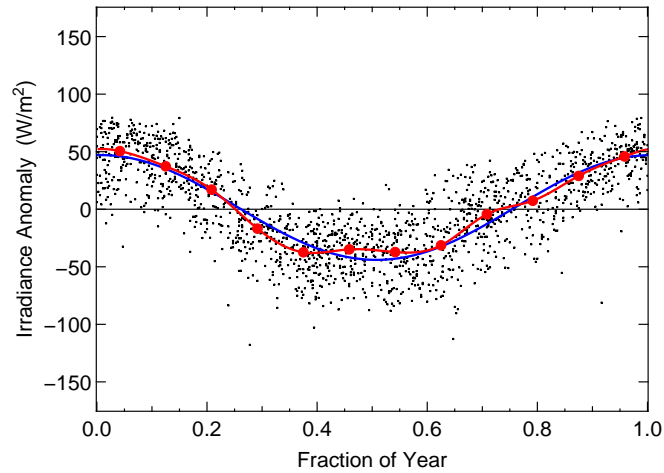}}
\caption{Average seasonal cycle in the terrestrial irradiance at MLO
  and airmass 2 computed from the irradiance data in the years 2000-2009
  (small black squares) and after removing a linear long-term
  trend. The filled red circles are the monthly medians while the red
  line shows a fit of cubic splines with periodic boundary conditions
  through the monthly medians. The blue line indicates the expected
  seasonal irradiance variations due to Earth's orbital
  motion.}\label{f:seasons}
\end{figure}

For the MLO data the years from 2000 to 2009 have been chosen to
compute the average seasonal cycle (indicated by the grey shading in
Figure~\ref{f:mlotimecorr}a) as this period of time is only weakly
affected by volcanic aerosols. Furthermore, a linear trend with time
has been subtracted before computing the monthly median irradiance
values to remove any long-term changes in the irradiance baseline (be
it from solar activity changes, trends in volcanic or other aerosols,
or instrumental drift). The resulting cycle of seasonal irradiance
anomalies is shown in Figure~\ref{f:seasons}.

The irradiance data with the seasonal cycle anomaly removed are shown
in Figure~\ref{f:mlotimecorr}b. The irradiance variations through the
year are clearly dominated by the changes in the distance to the Sun
on Earth's elliptical orbit. In addition there are minor, but
statistically insignificant, seasonal anomalies, in particular in
spring and summer.

Note that there is quite a bit of interannual variation in the
seasonal irradiance cycle; therefore, a perfect correction of the effect
cannot be expected. Comparing the irradiance time series with the
seasonal cycle subtracted shown in Figure~\ref{f:mlotimecorr}b to the
raw irradiance data shows, however, that the seasonal cycle is
effectively removed. In particular, the sequence in the irradiance
vs.\ sunspot number scatter plot shown in the right-hand panel of
Figure~\ref{f:mlotimecorr}b exhibits less scatter and visually appears
considerably better defined.

\subsubsection{Effects of Volcanic Aerosols}

The second atmospheric effect which has to be taken into account when
investigating the correlation between solar activity and terrestrial
insolation are the attenuating effects of volcanic aerosols. One way
of dealing with episodes of volcanic eruptions is to exclude these
periods from further analysis as has been done for the SAO terrestrial
irradiance data in \citeauthor{Feulner2011b} (2011a). This is not a
viable option for the frequent volcanic eruptions affecting the MLO
data, however. The expected irradiance variations due to solar
activity are below 1\% or $\Delta I \lesssim 10$~W~m$^{-2}$
corresponding to an attenuation by volcanic aerosols with an optical
depth of $\tau \lesssim 0.01$. Such a cut in aerosol optical
depth would mean that more than half of the MLO irradiance record
would have to be excluded from the analysis, and even then the
remaining record would suffer from inhomogeneous attenuation by
volcanic aerosols resulting in irradiance variations at least of the
order of the changes expected from solar variability.

A preferable option would be to correct for the effects of volcanic
eruptions using an independent dataset of the optical depth of
volcanic aerosols. For the analysis in this paper, the
\citet{Sato1993} compilation of global average optical depth
$\tau_{550}$ of volcanic aerosols at wavelength $\lambda = 550$~nm is
used for the years 1958--1993, extended with the \citet{Solomon2011}
data until March 2010 and with a constant value $\tau_{550} = 0.007$
until the end of 2010. The irradiance $I_\mathrm{sv}$ corrected for
the seasonal cycle and the effects of volcanic aerosols is then
calculated from the irradiance $I_\mathrm{s}$, which is corrected for
the seasonal cycle only, using $I_\mathrm{sv} = I_\mathrm{s} \exp (X
\, c_X \, \tau_{550})$. In this formula, $X$ denotes the airmass, and
$c_X$ is a correction factor converting the optical depth at 550~nm to
the optical depth integrated over the whole visible solar
spectrum. The correction factors $c_X$ for the four airmass values are
determined empirically, finding $c_2 = 0.50 \pm 0.01$, $c_3 = 0.45 \pm
0.01$, $c_4 = 0.42 \pm 0.01$ and $c_5 = 0.40 \pm 0.01$.

The time series and scatter plot for the irradiance $I_\mathrm{sv}$
corrected for seasons and volcanic aerosols is shown in
Figure~\ref{f:mlotimecorr}c. Overall the correction seems to work well
for the late stages of volcanic eruptions. There are, however,
considerable residuals during times of rapidly changing volcanic
aerosol load, in particular for the Pinatubo and El~Chich\'on
eruptions. This is not surprising since the global average aerosol
optical depth may not be fully representative for the local aerosol
load above Mauna Loa, in particular in the period following the
eruptions. Furthermore, the \citet{Sato1993} compilation provides
monthly values which are not able to trace the fast changes visible in
the daily irradiance record. Two-year periods after the four strongest
eruptions (Agung, Fernadina, El~Chich\'on, and Pinatubo) will be
masked out for the analysis of the correlation between terrestrial
irradiance and sunspot number presented below. Note that there is also
a dip in the thus corrected irradiance around the year 1967,
corresponding to a decrease in volcanic aerosol optical depth in the
\citet{Sato1993} record. As the origin of this period of lower
irradiance remains unclear, it will not be excluded from the analysis.

\subsubsection{Long-Term Trend}

Looking at the scatter plot of irradiance values corrected for the
seasonal cycle and volcanic aerosols with sunspot number shown in the
right-hand panel of Figure~\ref{f:mlotimecorr}c one can indeed see a
marked increase of terrestrial solar irradiance with sunspot
number. At least part of this trend is clearly driven by what appears
to be a separate sequence of measurements at high irradiance ($I
\simeq 1100$~W~m$^{-2}$) at relatively high sunspot numbers ($100
\lesssim R \lesssim 300$).

A comparison with the irradiance time series shows that the vast
majority of these measurements are from the very first years of the
record (years 1958--1961) which, according to Figure~\ref{f:mlotime},
are unaffected by volcanic aerosols and coincide with the particularly
strong maximum of solar cycle~19.

These observations are important since the irradiance data shown in
Figure~\ref{f:mlotimecorr}c exhibit a slowly decreasing trend over
time. This general decrease in irradiance can be clearly seen during
the last decade, for example, or by comparing the beginning of the
record and the end. This trend in atmospheric transmission has been
noted before and attributed to changes in the atmospheric background
aerosol load \citep{Solomon2011}.

It will be shown below that this long-term trend can be detected in
the data spanning almost five solar cycles. Note, for example, that
the decrease can be seen over the entire solar cycle 23 with a
constant slope despite large variations in sunspot number. The trend
can thus be regarded as independent of solar activity and has to be
subtracted to ensure an accurate analysis of the correlation between
solar activity and terrestrial irradiance. To this end a linear trend
$I_\mathrm{t} \, (t) \, = \, I_0 + a_I \, t$ is approximated to the
entire record of irradiance data corrected for the seasonal cycle and
volcanic aerosols, but excluding the two-year time intervals after the
El~Chich\'on and Pinatubo eruptions. Furthermore, the data used to fit
the linear trend are restricted to the range $0 < R < 200$ to ensure
that the fit of the long-term trend is not driven by changes in solar
activity between the strong solar maximum of cycle 19 and the
considerably less pronounced maximum of cycle 23. Fitting a linear
trend using these data yields a slope of $a_I = -0.38 \pm
0.02$~W~m$^{-2}$~yr$^{-1}$. This is very similar to the trend found in
\citet{Solomon2011}. As will be shown below, a multivariate regression
has been performed on $I_\mathrm{sv}$ to simultaneously analyse its
linear trends with sunspot number and time as an additional test
whether the trend in time reported here is indeed independent of solar
activity, finding a very similar value for the long-term decrease in
terrestrial irradiance with time.

Thus there is indeed a significant long-term decrease in terrestrial
irradiance which is independent of solar activity and therefore has to
be subtracted from the data before analysing any correlation of
irradiance with sunspot number. The resulting irradiance time series
of $I_\mathrm{svt} (t) \, = \, I_\mathrm{sv} (t) - I_\mathrm{t} (t)$
and its scatter plot are shown in Figure~\ref{f:mlotimecorr}d. It
should be noted that the conspicuous second sequence at large
irradiance values in the scatter plot has now disappeared and that the
overall correlation shows a considerably smaller slope than before the
subtraction of the linear long-term trend.

\section{Correlation of Terrestrial Irradiance with Sunspot Number}
\label{s:correlation}

After applying the corrections for the seasonal cycle, for the
attenuating effects of volcanic aerosols, and for the linear long-term
trend, the MLO terrestrial irradiance data at morning airmass 2 can
now be correlated with the sunspot number to investigate changes of
terrestrial irradiance with solar activity.

\begin{figure}[ht!]
\centerline{\includegraphics[width=0.7\textwidth]{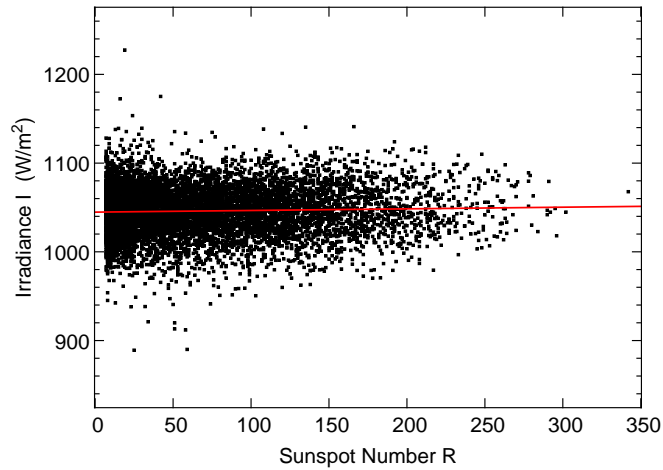}}
\caption{Scatter plot of MLO terrestrial irradiance vs.\ sunspot
  number (black squares). The irradiance data are corrected for the
  seasonal cycle, volcanic aerosols, and a long-term decrease in
  atmospheric transmission independent of solar activity. The red line
  shows a linear fit to the correlation.}\label{f:correlation}
\end{figure}

The corrected irradiance data $I_\mathrm{svt}$ as a function of
sunspot number $R$ are presented in
Figure~\ref{f:correlation}. Fitting a line to this correlations yields
a slope 

\begin{equation}
  \mathrm{d} I_\mathrm{svt} / \mathrm{d}R \, = \, + (0.019 \pm 0.007)
  \, \mathrm{W\,m}^{-2} . 
\end{equation}
The error has been computed from 10\,000 bootstrapping simulations
where the sample is first duplicated and added to the original sample,
then modified by random errors, before half of the enlarged sample is
randomly selected to perform the linear regression (see also
\citeauthor{Feulner2011b}, 2011a).  A random error of 0.5\% is assumed
for the individual MLO irradiance measurements \citep{Dutton2001}
during these bootstrapping simulations.

In the analysis so far, the long-term trend with time $t$ of the
irradiance $I_\mathrm{sv}$ corrected for the seasonal cycles and the
effects of volcanic aerosols has been removed before correlating with
sunspot number $R$. While this is highly illustrative, it is
statistically more appropriate to perform a simultaneous multivariate
linear regression of $I_\mathrm{sv}$ with respect to trends with $t$
and $R$. Thus one can also independently test whether the linear trend
in time reported in Section~\ref{s:corrections} is indeed independent
of solar activity as traced by the sunspot number $R$. This test
yields a linear trend of terrestrial irradiance with time of $\partial
I_\mathrm{sv}/\partial t = -0.45 \pm 0.02$~W~m$^{-2}$~yr$^{-1}$, which
is slightly larger than the value reported above, but compatible
within the errors. The slope of the correlation with sunspot number
derived in this way is

\begin{equation}
  \partial I_\mathrm{sv} / \partial R \, = \, + (0.015 \pm 0.006) \,
  \mathrm{W\,m}^{-2} ,
\end{equation}
again in excellent agreement with the value from Equation~(1).

Thus a simultaneous regression of a linear long-term trend with time
and a linear trend with sunspot number to the terrestrial irradiance
data not yet corrected for the linear trend yields results which are
very similar to the ones described above. This is important because if
the long-term trend in terrestrial irradiance had been caused by
changes in solar activity, this long-term trend would be reflected in
the sunspot numbers as well, and the regression would have yielded an
insignificant trend with time. This clearly shows that the linear
long-term trend described above is indeed not caused by solar
activity.

As a test whether this result depends on the choice of sunspot number
as indicator of solar activity, the simultaneous regression has been
repeated for annual means of the open solar flux $F_\mathrm{S}$
\citep{Lockwood2009b}, finding consistent values for the trends,
albeit with larger errors ($\partial I_\mathrm{sv}/\partial t = -0.47
\pm 0.15$~W~m$^{-2}$~yr$^{-1}$ and $\partial I_\mathrm{sv}/\partial
F_\mathrm{S} = (1.4 \pm 2.6) \times 10^{-14}$~W~m$^{-2}$~Wb$^{-1}$).

In summary, a positive trend of the corrected terrestrial irradiance
data with sunspot number can be found in the MLO data which is
significant on the $\simeq 2\sigma$ level. The magnitude of these
changes of solar irradiance on the ground with respect to changes in
irradiance on top of the atmosphere and previous studies on
correlations of terrestrial irradiance with solar activity will be
discussed in Section~\ref{s:disc}.

\section{Correlation of Atmospheric Transmission with Sunspot Number}
\label{s:atf}

\begin{figure}[ht!]
\centerline{\includegraphics[height=0.95\textwidth,angle=270]{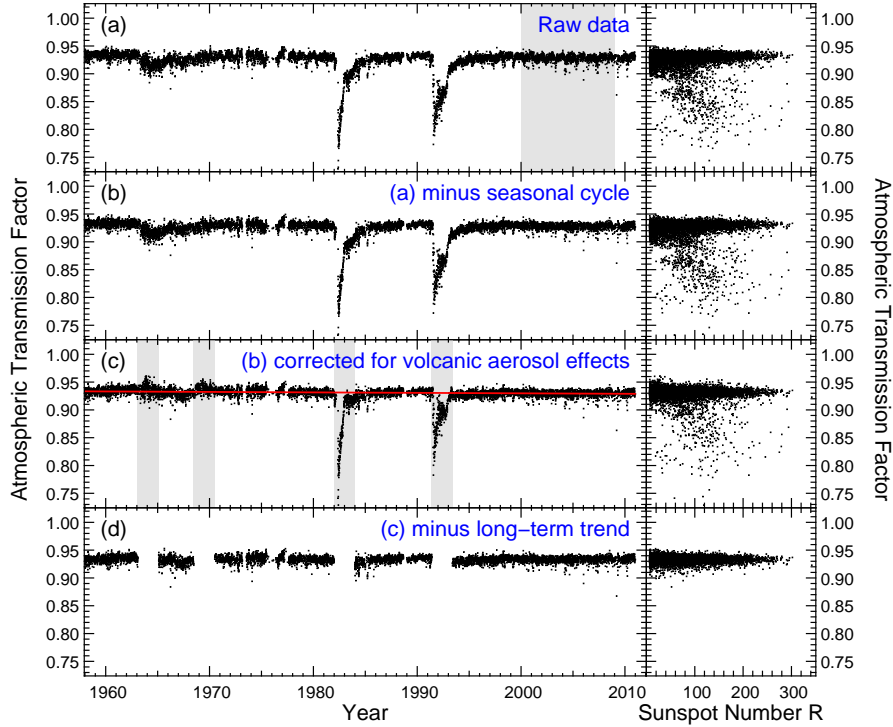}}
\caption{Same as Figure~\ref{f:mlotimecorr}, but for the atmospheric
  transmission factor (ATF) and the corrections applied to this
  quantity.}\label{f:mlotimeatf}
\end{figure}

The analysis above investigated the correlation of terrestrial
insolation with sunspot number to investigate the possibility of a so-far
unknown amplification effect produced by changes in clear-sky
atmospheric transmission. Alternatively, one can directly study the
atmospheric transmission itself. The advantage of this approach is
that one can make use of the rationing technique which eliminates both
the variations in TSI and any instrumental calibration differences
\citep{Ellis1971}. In this method, atmospheric transmission is
analysed in terms of the atmospheric transmission factor (ATF) defined
as follows:

\begin{equation}
  \mathrm{ATF} \: = \: \frac{1}{3} \left( \frac{I_3}{I_2} \, + \,
    \frac{I_4}{I_3} \, + \, \frac{I_5}{I_4} \right) ,
\end{equation}
where $I_X$ are the morning irradiance values at the different airmass
values $X$ of 2, 3, 4, and 5. The ATF for the Mauna Loa data is shown
in Figure~\ref{f:mlotimeatf}a.

Before correlating the ATF with sunspot number $R$, the same
corrections have to be applied as for the irradiance data (see
Section~\ref{s:corrections}). The seasonal cycle in the ATF is
computed as for the irradiance, but note that the dominant effect
caused by the orbital motion of the Earth is already taken care of by
the rationing technique. The remaining corrections for the effects of
volcanic aerosols and the long-term trend (with a best-fitting value
of $\mathrm{d(ATF)} / \mathrm{d} t = -(8.2 \pm 0.4) \times
10^{-5}$~yr$^{-1}$) also follow the procedure discussed in detail for
the irradiance. The stepwise corrections and their effect both on the
time series and for the correlation with sunspot number are
illustrated in Figure~\ref{f:mlotimeatf}.

\begin{figure}[ht!]
\centerline{\includegraphics[width=0.7\textwidth]{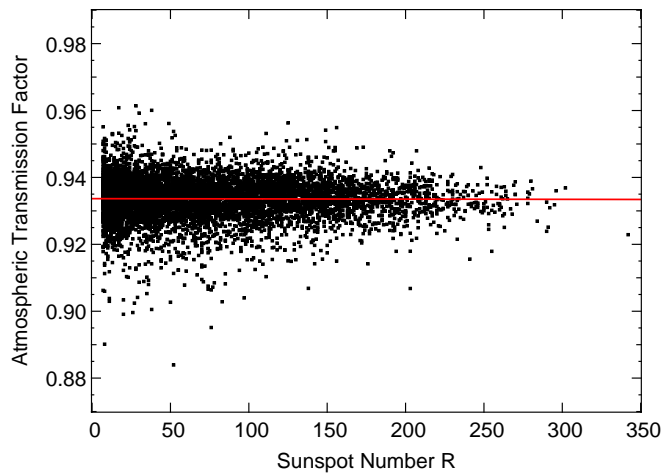}}
\caption{Scatter plot of MLO atmospheric transmission factor (ATF)
  vs.\ sunspot number (black squares). The ATF data are corrected for
  the seasonal cycle, volcanic aerosols, and a long-term decrease in
  atmospheric transmission independent of solar activity. The red line
  shows a linear fit to the correlation.}\label{f:atfcorrelation}
\end{figure}

The correlation between the corrected atmospheric transmission factor
ATF and sunspot number $R$ is shown in
Figure~\ref{f:atfcorrelation}. A linear fit to this correlation yields

\begin{equation}
  \mathrm{d(ATF)} / \mathrm{d}R \, = \, - (0.7 \pm 1.4) \times 10^{-6} .
\end{equation}

As for the irradiance data, an improved analysis is performed in which
both the long-term trend of the ATF with time and its correlation with
sunspot number are fit simultaneously. This analysis yields a value of
$\partial \mathrm{(ATF)} / \partial t = -(10.5\pm 0.4) \times
10^{-5}$~yr$^{-1}$ for the long-term trend, in good agreement with the
value derived above. The fact that the simultaneous linear regression
of the ATF finds a significant long-term trend with time again
confirms that this long-term trend is independent of solar
activity. The linear correlation of the ATF with sunspot number has a
slope of $\partial \mathrm{(ATF)} / \partial R = -(2.0 \pm 1.3) \times
10^{-6}$, again in agreement with Equation~(4). Thus no significant
change of atmospheric transmission with sunspot number can be detected
in the Mauna Loa data.

To test for a possible dependence of this result on the choice of
solar-activity indicator, the bivariate regression has been repeated
replacing the sunspot number $R$ by the open solar flux $F_\mathrm{S}$
(using annual averages of the data as for the analysis of the
irradiance data). Again, the trends derived for the open solar flux
are in good agreement with the trends computed using sunspot numbers,
although the errors are larger ($\partial \mathrm{(ATF)}/\partial t =
(-10.7 \pm 2.7)$~yr$^{-1}$ and $\partial \mathrm{(ATF)}/\partial
F_\mathrm{S} = (-1.6 \pm 4.8) \times 10^{-18}$~Wb$^{-1}$).

\section{Discussion}
\label{s:disc}

The analysis in Section~\ref{s:correlation} shows that the corrected
terrestrial insolation $I_\mathrm{sv}$ linearly increases with sunspot
number $R$ with a slope $\partial I_\mathrm{sv} / \partial R \, = \, +
(0.015 \pm 0.006)$~W~m$^{-2}$. With typical differences in sunspot
number of $\Delta R \approx 150$ between solar maxima and minima, this
translates into irradiance variations of $\Delta I \approx 2 \pm
1$~W~m$^{-2}$ or $\sim 0.2 \pm 0.1$\% over the solar cycle. This is of
the same order of magnitude as the variations in total solar
irradiance (TSI) observed from satellites
\citep[\textit{e.g.},][]{Frohlich2004, Gray2010} which suggests that
clear-sky terrestrial irradiance variations with solar activity are
not strongly amplified by any hitherto unknown atmospheric feedback
processes.

The conflicting results in \citet{Hempelmann2012} who find a one order
of magnitude larger variation in terrestrial irradiance with solar
activity ($dI/dR \, = \, (0.10-0.18)$~W~m$^{-2}$ depending on the way
the effects of volcanic eruptions are accounted for) can be understood
from the fact that the authors have not corrected the terrestrial
irradiance data for the slow long-term decrease in atmospheric
transmission not related to changes in solar activity. The combination
of large irradiance values at the beginning of the record with large
sunspot numbers during the strong maximum of solar cycle 19 then
results in a spuriously large slope of the irradiance vs.\ sunspot
number correlation. In addition, excluding periods of up to three
years after volcanic eruptions as done in \citet{Hempelmann2012} is
insufficient as attenuation effects at the percent level are visible
in the data even after this time, resulting in changes of at least the
order of magnitude of the changes related to solar variability (see
the discussion in Section~\ref{s:corrections}).

The lack of evidence for any unknown effect amplifying changes of
terrestrial irradiance with solar activity in the MLO data is
confirmed by an analysis of the correlation of the atmospheric
transmission factor with sunspot number which failed to find a
significant change of clear-sky atmospheric transmission with solar
activity.

\section{Conclusions}
\label{s:concl}

In this paper, the Mauna Loa Observatory (MLO) direct-beam solar
irradiance record has been analysed to investigate the correlation
between solar activity and terrestrial insolation. After correcting
for the seasonal cycle, the attenuating effects of volcanic aerosols,
and a long-term decrease in atmospheric transmission independent of
solar activity, clear-sky terrestrial irradiance is found to vary by
$\Delta I \approx 0.2 \pm 0.1$\% between maxima and minima of the
11-year solar activity cycle. These variations are of the same order
of magnitude as the changes in total solar irradiance on top of the
atmosphere over the solar activity cycle. An investigation of the
atmospheric transmission shows that there is no significant trend of
clear-sky atmospheric transmission with sunspot number. Thus there is
no evidence for an unknown amplification effect of solar activity in
terms of strong changes in clear-sky atmospheric transmission.

%

%

%

%
\begin{acks}
  I would like to thank Ellsworth G.\ Dutton for providing me with the
  latest version of the Mauna Loa Observatory terrestrial solar
  irradiance data and for many helpful comments on these data. I am
  grateful to the anonymous referee for a thorough review. This
  research has made use of NASA's Astrophysics Data System
  Bibliographic Services.
\end{acks}

%
%

%
%
%
%

\end{article} 
\end{document}